\documentclass[a4paper,onecolumn,12pt]{article}
\usepackage{graphicx}
\usepackage{hhline}
\usepackage{mathrsfs}
\usepackage{amssymb}
\usepackage{setspace}

\begin{document}
\Large\textbf{\textrm{\hspace{-1.2em} Numerical Analysis of Relativistic Boltzmann-kinetic Equations to Solve Relativistic Shock Layer Problems}}\\ \\
\large\textmd{Ryosuke Yano}\\
\small\textit{Department of Advanced Energy, University of Tokyo, 5-1-5 Kashiwanoha, Kashiwa, Chiba 277-8561, Japan \\
Email: yano@daedalus.k.u-tokyo.ac.jp}\\ \\
\large\textmd{Kojiro Suzuki}\\
\small\textit{Department of Advanced Energy, University of Tokyo, 5-1-5 Kashiwanoha, Kashiwa, Chiba 277-8561, Japan \\
Email: kjsuzuki@k.u-tokyo.ac.jp, Main telephone number: +81-4-7136-3828,\\
Main fax number: +81-4-7136-3828}
\\ \\
\large\textmd{Hisayasu Kuroda}\\
\small\textit{Super Computing Division, Information Technology Center, University of Tokyo, 2-11-16 Yayoi, Bunkyo, Tokyo, Japan \\
Email: kuroda@pi.cc.u  -tokyo.ac.jp}\\ \\
\textsf\textbf{ABSTRACT}\\\\
\textrm{The relativistic shock layer problem was numerically analyzed by using two relativistic Boltzmann-kinetic equations. One is Marle model, and the other is Anderson-Witting model. As with Marle model, the temperature of the gain term was determined from its relation with the dynamic pressure in the framework of 14-moments theory. From numerical results of the relativistic shock layer problem, behaviors of projected moments in the nonequilibrium region were clarified. Profiles of the heat flux given by Marle model and Anderson-Witting model were quite adverse to the profile of the heat flux approximated by Navier-Stokes-Fourier law. On the other hand, profiles of the heat flux given by Marle model and Anderson-Witting model were similar to the profile approximated by Navier-Stokes-Fourier law.
Additionally we discuss the differences between Anderson-Witting model and Marle model by focusing on the fact that the relaxational rate of the distribution function depends on both flow velocity and molecular velocity for Anderson-Witting model, while it depends only on the molecular velocity for Marle model.}
\newpage
\hspace{-1.4em}\textsf{\textbf{I. INTRODUCTION}}\\\\
Advances in elementary particle physics has made the study of the flow field of relativistic particles an important issue. The study of relativistic Boltzmann equations is particularly important for understanding the fundamental properties of nonequilibrium relativistic gases. Despite this importance, numerical analysis of relativistic Boltzmann equations \cite{Cercignani} has not yet been reported. Fortunately, two relativistic kinetic equations can be analyzed numerically due to their simplified collision kernels. One is the Anderson-Witting model \cite{Anderson}, and the other is the Marle model \cite{Marle}. Both models are written in BGK (Bhatnagar-Gross-Krook) \cite{BGK} form. Assumptions in the construction of the numerical scheme to solve either the Anderson-Witting or Marle models include an inertial frame and no degeneracy of the particles. Projected moments \cite{Str} are calculated using a standard method, namely, the Eckart decomposition \cite{Eckart}.\\

In our analysis of the Marle model, we focused on the dynamic pressure, which is a characteristic quantity for relativistic gases. In the Marle model, the dynamic pressure is derived from the difference in temperatures of the gain and loss terms. The Anderson-Witting model avoids this inequality by using orthogonality of the four velocities of the flow by Landau-Lifshitz \cite{Landau} to the nonequilibrium terms. However a concrete formulation of the temperature of the gain term has not yet been included in the Marle model. In our analysis, we therefore related the temperature of the gain term with the dynamic pressure in the framework of the N.S.F. (Navier-Stokes-Fourier) law derived from 14-moment theory \cite{Cercignani}\cite{Stewart} for the Marle model. Provided that we equate the temperatures of the gain and the loss terms in the Marle model, the bulk viscosity always becomes negative.\\

To quantitatively examine the behavior of the Anderson-Witting and Marle models, we used a numerical method to solve the relativistic, steady-state shock-layer problem in the absolute standard of rest {\sf The meaning of "absolute standard of rest is not clear". Please explain what you mean by this phrase.}, which is a hypothetical inertial frame. The heat flux and the dynamic pressure were determined from the simulations. The shock layer problem is suitable to investigate nonequilibrium gas dynamics in both the shock structure and in the boundary layer. \\

Finally, we discuss differences between the Anderson-Witting and Marle models by comparing their flow and molecular-velocity dependent relaxation of the distribution function {\sf Do you mean "velocity distribution function"?}. Significant differences in relaxation of their distribution functions is observed in the negative tail structures {\sf The meaning of "negative tail structure" is not clear. Please explain what you mean.} of their distribution functions at the shock's rising edge. Throughout this paper, the molecular potential is represented as a hard sphere molecule.\\

This paper is organized as follows. In Section II we review the relativistic Boltzmann equation and its relation to the projected moments with Eckart decomposition. In Section III we describe the Anderson-Witting and Marle models, and describe a derivation of the temperature of the gain term from the N.S.F. law. In Section IV we describe a numerical method for solving the Anderson-Witting and Marle models. In Section V we describe an application of this method to solve the shock layer problem. \\\\
\textsf{\textbf{II. RELATIVISTIC BOLTZMANN EQUATION}}\\\\
The relativistic Boltzmann equation is written as \cite{Cercignani}
\begin{eqnarray}
p^{\alpha}\frac{\partial f}{\partial x^\alpha}&=&Q(f,f) \nonumber \\
&=&\int_{\mathscr{R}^3} \left(f_{\ast}^\prime f^\prime -f_{\ast}f \right)F\sigma d\Omega \frac{d^3 \mathbf{p}_\ast}{p_{\ast 0}},
\end{eqnarray}
where $x^\alpha$ represents the four-dimensional coordinates, $p^\alpha$ is a four-dimensional momentum vector, $f$ is a distribution function defined by $f=f\left(t,x^i,p^i\right) \ (i=1,2,3)$, and $F$ is the Lorentz invariant flux. In eq. (1), terms with prime indicate conditions after collisions and $\mathscr{R}^3$ is the momentum space stretched by $\left\{ \mathscr{R}^3 | (-\infty,-\infty,-\infty) \le (p^1,p^2,p^3) \le (\infty,\infty,\infty)\right\}$. $x^\alpha$, $p^\alpha$ and $F$ are given by
\begin{eqnarray}
x^\alpha&=&(ct,x^1,x^2,x^3), \\
p^\alpha&=&m\gamma({v})\left(c,v^1,v^2,v^3\right),\\
F&=&\frac{p^0p_{\ast}^0}{c} g_{\o}=\frac{p^0p_{\ast}^0}{c} \sqrt{\left(\mathbf{v}-\mathbf{v}_{\ast} \right)^2-\frac{1}{c^2}\left(\mathbf{v}\times\mathbf{v}_{\ast}\right)^2} \nonumber \\
&=&\sqrt{\left(p_{\ast}^\alpha p_{\alpha}\right)^2-m^4c^4}.
\end{eqnarray}
In eq. (3), $\gamma(v)$ is the Lorentz factor, which is given by $\gamma(v)={1}/{\sqrt{1-v^2/c^2}}$. $c$ is the speed of light and $v^i$ is the $i$th component of the particle velocity for $i \ (=1,2,3)$. In eqs. (3) and (4), $m$ is the molecular mass. In eq. (4), $g_{\o}$ is M\o ller's relative velocity. In eq. (1), $\sigma$ is the differential cross section and $\Omega$ is the solid angle. In eqs. (1) and (4), terms with an asterisk subscript belong to the collision partner. Rewriting eq. (1) in Lorentz variant form yields
\begin{eqnarray}
\frac{\partial f}{\partial t}+v^i\frac{\partial f}{\partial x^i}=\int_{\mathscr{R}^3} \left(f_{\ast}^\prime f^\prime -f_{\ast}f \right)g_\phi \sigma d\Omega d^3 \mathbf{p}_{\ast}.
\end{eqnarray}
Multiplying both sides of eq. (1) by $p^\alpha$ and $p^{\alpha}p^\beta$ and integrating in momentum space, we obtain conservation equations in terms of $N^\alpha=\int_{\mathscr{R}^3} cp^\alpha f \frac{d^3 \mathbf{p}}{p^0}$ and $T^{\alpha\beta}=\int_{\mathscr{R}^3} c p^\alpha p^\beta f \frac{d^3 \mathbf{p}}{p^0}$ as \newline
Mass conservation:
\begin{eqnarray}
&&\partial_\alpha N^\alpha \nonumber \\
&&=c\int_{\mathscr{R}^3} p^\alpha \frac{\partial f}{\partial x ^\alpha}\frac{d^3 \mathbf{p}}{p^0} \nonumber \\
&&=\frac{1}{2}\int_{\mathscr{R}^3 \times \mathscr{R}^3}  \left(c+c-c-c\right)ff_\ast F \sigma d \Omega \frac{d^3 \mathbf{p}_\ast}{p^0_\ast} \frac{d^3\mathbf{p}}{p^0} \nonumber \\
&&=0.
\end{eqnarray}
Momentum-Energy conservation:
\begin{eqnarray}
&&\partial_\alpha T^{\alpha\beta} \nonumber \\
&&=c\int_{\mathscr{R}^3}  p^\alpha p^\beta \frac{\partial f}{\partial x ^\alpha}\frac{d^3 \mathbf{p}}{p^0} \nonumber \\
&&=\frac{1}{2}\int_{\mathscr{R}^3 \times \mathscr{R}^3}\left({p_\ast^\beta}^\prime+{p^\beta}^\prime-p_\ast^\beta- p^\beta \right)ff_\ast F \sigma d \Omega \frac{d^3 \mathbf{p}_\ast}{p^0_\ast} \frac{d^3\mathbf{p}}{p^0}\nonumber \\
&&=0.
\end{eqnarray}
According to Eckart \cite{Eckart}, $N^\alpha$ and $T^{\alpha\beta}$ can be decomposed to
\begin{eqnarray}
N^\alpha&=&c\int_{\mathscr{R}^3} p^\alpha \frac{d^3\mathbf{p}}{p^0} =nU^\alpha,\\
T^{\alpha\beta}&=&c\int_{\mathscr{R}^3} p^\alpha p^\beta \frac{d^3\mathbf{p}}{p^0} \nonumber \\
&=& p^{<\alpha\beta>}-\left(p+\varpi\right)\Delta^{\alpha\beta} \nonumber \\
&& +\frac{1}{c^2}\left(U^\alpha q^\beta+U^\beta q^\alpha \right)+\frac{en}{c^2}U^\alpha U^\beta,
\end{eqnarray}
where $n$ is the number density, $p^{<\alpha\beta>}$ is the shear stress, $p$ is the isotropic pressure, $\varpi$ is the dynamic pressure, $q^{\alpha}$ is the heat flux, $e$ is the energy density, and $U^{\alpha}$ is the four-dimensional velocity field of the flow given by
\begin{eqnarray}
U^{\alpha}=\gamma(u)\left(c,u^i\right),
\end{eqnarray}
where $u^i$ is the $i$th component of the flow velocity. $\Delta^{\alpha\beta}$ in eq. (9) is the projector defined by
\begin{eqnarray}
\Delta^{\alpha\beta}=\eta^{\alpha\beta}-\frac{1}{c^2}{U^\alpha U^\beta},
\end{eqnarray}
where $\eta^{\alpha \beta}$ is given by
\begin{eqnarray}
\eta^{\alpha\beta}=\eta_{\alpha\beta}=\pmatrix{ +1 & 0 & 0 & 0 \cr 0 & -1 & 0 & 0 \cr 0 & 0 & -1 & 0 \cr 0 & 0 & 0 & -1 \cr }.
\end{eqnarray}
Projected moments are obtained as \cite{Eckart}
\begin{eqnarray}
n&=&\frac{1}{c^2}N^\alpha U_{\alpha},\\
p^{<\alpha\beta>}&=&\left(\Delta_{\gamma}^{\alpha}\Delta_{\delta}^{\beta}-\frac{1}{3}\Delta^{\alpha\beta}\Delta_{\gamma\delta}\right)T^{\delta\gamma}\\
p+\varpi &=&-\frac{1}{3}\Delta_{\alpha\beta}T^{\alpha\beta},\\
q^\alpha&=&\Delta_{\gamma}^{\alpha}U_{\beta}T^{\beta\gamma},\\
e&=&\frac{1}{nc^2}U_\alpha T^{\alpha\beta}U_{\beta},
\end{eqnarray}
Projected moments $n,u^i,p^{<\alpha\beta>},q^\alpha$ and $\varpi$ can be reduced from eqs. (14) and (16) to 14 projected moments $n,u^i,p^{ij},q^i,\varpi, \ (i,j=1,2,3)$ as
\begin{eqnarray}
p^{<\alpha\beta>}U_\alpha&=&0,\\
q^\alpha U_\alpha&=&0,
\end{eqnarray}
\\
 Conservative equations in eqs. (6) and (7) yield balance equations for $n$ (mass), $U^\alpha$ (momentum), and $e$ (energy)\cite{Cercignani} as
\begin{eqnarray}
&&Dn+n \nabla^\alpha U_\alpha=0, \\
&&\frac{nh_E}{c^2} DU^\alpha=\nabla^\alpha \left(p+\varpi \right)-\nabla_\beta p^{<\alpha \beta>} \nonumber \\
&& +\frac{1}{c^2}\left(p^{<\alpha\beta>}DU_\beta-\varpi DU^\alpha-Dq^\alpha-q^\alpha \nabla_\beta U^\beta-q^\beta \nabla_\beta U^\alpha-\frac{1}{c^2}U^\alpha q^\beta DU_\beta-U^\alpha p^{<\beta\gamma>}\nabla_\beta U_\gamma \right) \nonumber \\ \\
&& nDe=-\left(p+\varpi \right)\nabla_\alpha U^\alpha+p^{<\alpha\beta>}\nabla_\beta U_\alpha-\nabla_\alpha q^\alpha+\frac{2}{c^2}q^\alpha DU_\alpha
\end{eqnarray}
where $D$, $\nabla^\alpha$, and the enthalpy per particle $h_E$ are defined by
\begin{eqnarray}
&&D\equiv U^\alpha \partial_\alpha, \\
&&\nabla^\alpha=\left(\eta^{\alpha\beta}-\frac{1}{c^2}U^\alpha U^\beta\right) \partial_\beta=\Delta^{\alpha\beta} \partial_\beta, \\
&&h_E=e+\frac{p}{n},
\end{eqnarray}
where $\partial_\alpha\equiv \frac{\partial}{\partial x^\alpha}$. Chapmann-Enskog expansion indicates that $\varpi$, $p^{<\alpha\beta>}$ and $q^\alpha$ are approximated by the product of temporal-spatial gradients of projected moments and transport coefficients, the bulk viscosity $\eta$, the viscosity coefficient $\mu$, and the thermal conductivity $\lambda$ as follows \cite{Cercignani},
\begin{eqnarray}
&&\varpi=-\eta \nabla_\alpha U^\alpha, \\
&&p^{<\alpha\beta>}=2\mu\left[\frac{1}{2}\left(\Delta_\gamma^\alpha\Delta_\delta^\beta+\Delta_\delta^\alpha\Delta_\gamma^\beta\right)-\frac{1}{3}\Delta^{\alpha\beta}\Delta_{\gamma\delta} \right] \nabla^\gamma U^\delta \nonumber \\
\\
&&q^\alpha=\lambda\left(\nabla^\alpha \theta -\frac{\theta}{nh_E}\nabla^\alpha p\right)
\end{eqnarray}
where $\theta$ is the gas temperature.
\\\\
\textsf{\textbf{III. RELATIVISTIC KINETIC EQUATION}}\\\\
In this section, we describe two kinetic equations: the Anderson-Witting and Marle models.\\\\
\textbf{Anderson-Witting model}\\\\
The Anderson-Witting model is given by \cite{Anderson}
\begin{eqnarray}
p^\alpha \frac{\partial f}{\partial x^\alpha}=\frac{U_L^\alpha p_{\alpha}}{c^2 \tau}\left(f^{(0)}-f\right),
\end{eqnarray}
where $f^{(0)}$ is an equilibrium function called the Maxwell-J\"uttner function, and can be defined as
\begin{eqnarray}
f^{(0)}(n,\theta_E,u)=\frac{n}{4\pi m^2 ck \theta_E K_2(\zeta_E)}e^{-\frac{U^\alpha p_\alpha}{k\theta_E}}.
\end{eqnarray}
where $\zeta_E$ is given by $\zeta_E=\frac{mc^2}{k\theta_E}$, $k$ is the Boltzmann constant, and $\theta_E$ is the temperature used in the equilibrium function $f^{(0)}$. $K_n$ is the $n_{th}$ order modified Bessel function and $U_L^\alpha$ is the four-dimensional velocity of the flow defined by Landau-Lifshitz \cite{Landau} and written as
\begin{eqnarray}
U_L^\alpha=U^\alpha+\frac{q^\alpha}{ne+p}.
\end{eqnarray}
$\tau$ in eqs. (29) is defined for the hard sphere molecule as \cite{Cercignani}
\begin{eqnarray}
\tau&=&\frac{1}{4n\pi \sigma v_s}, \\
v_s&=&\sqrt{\frac{\zeta^2+5G\zeta-G^2\zeta^2}{G\left(\zeta^2+5G \zeta-G^2\zeta^2-1 \right)}\frac{k\theta}{m}},\nonumber \\
\end{eqnarray}
where $v_s$ is the relativistic speed of sound and $G=K_3(\zeta)/K_2(\zeta)$.\\
For the conservation law for $N^\alpha$ and $T^{\alpha\beta}$ denoted by eqs. (6) and (7), the following constraints must be satisfied.
\begin{eqnarray}
&& c \int_{\mathscr{R}^3} p^\alpha U_{L\alpha}f \frac{d^3 \mathbf{p}}{p^0}=N^\alpha U_{L \alpha}=c\int_{\mathscr{R}^3} p^\alpha U_{L\alpha}f^{(0)} \frac{d^3 \mathbf{p}}{p^0}=N_E^\alpha U_{L\alpha},\\
&& c \int_{\mathscr{R}^3} p^\alpha p^\beta U_{L\alpha}f \frac{d^3 \mathbf{p}}{p^0}=T^{\alpha\beta} U_{L \alpha}=c\int_{\mathscr{R}^3} p^\alpha p^\beta  U_{L\alpha}f^{(0)} \frac{d^3 \mathbf{p}}{p^0}=T_E^{\alpha\beta} U_{L\alpha},
\end{eqnarray}
where subscript $E$ indicates quantities derived from the equilibrium distribution function $f^{(0)}$. These constraints are considered to be satisfied by the orthogonality of $U_L^\alpha$ to nonequilibrium terms in either $N^\alpha$ or $T^{\alpha\beta}$. Multiplying eq. (35) by $U_{L\beta}/(nc^2)$, we obtain \cite{Cercignani}\cite{Landau}
\begin{eqnarray}
e_E=e.
\end{eqnarray}
This relation in eq. (36) originates from the relation $e=mc^2\left(G(\zeta)-\frac{1}{\zeta} \right)$,
\begin{eqnarray}
\theta_E=\theta.
\end{eqnarray}
In the Anderson-Witting model $\theta_E$ is the temperature used in the equilibrium function $f^{(0)}$, which is equal to $\theta$, the temperature of the gas.\\\\
\textbf{Marle model} \\ \\
The Marle model \cite{Marle} is obtained by replacing $\frac{U_L^\alpha p_\alpha}{c^2}$ in eq. (29) by $m$. The Marle model can be written as
\begin{eqnarray}
p^\alpha \frac{\partial f}{\partial x^\alpha}=\frac{m}{\tau}\left(f^{(0)}-f\right),
\end{eqnarray}
where $\tau$ is defined in eq. (32).\\
By multiplying both sides of eq. (38) by $p^\beta p^\gamma$ and integrating in momentum space, we obtain
\begin{eqnarray}
\partial_\alpha T^{\alpha\beta\gamma}=\frac{m}{\tau}\left(T_E^{\beta\gamma}-T^{\beta\gamma}\right),
\end{eqnarray}
where $T^{\alpha\beta\gamma}=\int_{\mathscr{R}^3} p^\alpha p^\beta p^\gamma f \frac{d^3 \mathbf{p}}{p^0}$ and can be decomposed to yield \cite{Cercignani}
\begin{eqnarray}
T^{\alpha\beta\gamma}&=&\left(n C_1+C_2 \varpi \right)U^\alpha U^\beta U^\gamma+\frac{c^2}{6}\left(nm^2-nC_1-C_2 \varpi\right)\left(\eta^{\alpha\beta}U^\gamma+\eta^{\alpha\gamma}U^\beta+\eta^{\beta\gamma}U^\alpha \right) \nonumber \\
&&+C_3 \left(\eta^{\alpha\beta}q^\gamma+\eta^{\alpha\gamma}q^\beta+\eta^{\beta\gamma}q^\alpha\right)-\frac{6}{c^2}C_3\left(U^\alpha U^\beta q^\gamma +U^\alpha U^\gamma q^\beta +U^\beta U^\gamma q^\alpha \right) \nonumber \\
&&+C_4 \left(p^{<\alpha\beta>}U^\gamma+p^{<\alpha\gamma>}U^\beta+p^{<\beta\gamma>}U^\alpha\right).
\end{eqnarray}
$C_1,C_2,C_3$ and $C_4$ are functions of $\zeta$ shown in \cite{Cercignani}.\\
Multiplying both sides of eq. (39) by $U_\beta U_\gamma$ {\sf It is unclear what you are trying to say with "with eqs. (17) and (40)".} and eliminating terms with nonequilibrium projected moments, we obtain
\begin{eqnarray}
e_E-e=-\frac{\psi(\zeta_E)}{n} \nabla^\alpha U_\alpha,
\end{eqnarray}
where $\psi$ is defined in Appendix A. From eq. (41), the energy per particle $e$ is not conserved in the collision term in the Marle model.\\
Multiplying both sides of eq. (39) by $\Delta_{\beta\gamma}$ {\sf with eqs. (17) and (40)} and eliminating terms with nonequilibrium projected moments, we obtain
\begin{eqnarray}
\varpi=-\eta(\zeta_E) \nabla^\alpha U_\alpha=-\left(\hat\eta(\zeta_E)+\tilde \eta(\zeta_E)\right)\nabla^\alpha U_\alpha.
\end{eqnarray}
In eq. (42), $-\hat\eta(\zeta_E)\nabla^\alpha u_\alpha$ is the dynamic pressure derived from either $p_E-p$ or $e_E-e$ and $-\tilde \eta(\zeta_E) \nabla^\alpha U_\alpha$ is the dynamic pressure derived from the left hand side of eq. (39). If $\zeta=\zeta_E$, $-\hat\eta(\zeta_E)\nabla^\alpha u_\alpha = 0$, $\varpi=-\tilde\eta(\zeta)\nabla^\alpha U_\alpha$. $\tilde\eta(\zeta)$ and $\eta(\zeta_E)$ are given in Appendix A and are plotted in Fig. 1. Figure 1 shows that $\tilde\eta(\zeta)$ is negative for all ranges of $\zeta$ $(0 < \zeta)$.\\
From eqs. (40) and (41), $e_E$ is given by
\begin{eqnarray}
e_E=e+\frac{\psi(\zeta_E)}{n\eta(\zeta_E)} \varpi.
\end{eqnarray}
From the following approximate relation between $e$ and $p$ \cite{Cercignani},
\begin{eqnarray}
p_E-p=-\frac{n\left(e_E-e\right)}{1-5G_E\zeta_E-\zeta_E^2+G_E^2\zeta_E^2}=\frac{nk(e_E-e)}{C_v(\zeta_E)},
\end{eqnarray}
we obtain $\zeta_E$ or $\theta_E$ as
\begin{eqnarray}
&&\frac{1}{\zeta_E}=\frac{1}{\zeta}+\frac{k\psi(\zeta_E)}{nmc^2C_v(\zeta_E)\eta(\zeta_E)} \varpi,\\
&&\theta_E=\theta+\frac{\psi(\theta_E)}{nC_v(\theta_E)\eta(\theta_E)} \varpi,
\end{eqnarray}
where $C_v$ is the constant-volume specific heat.\\\\
\textsf{\textbf{IV. NUMERICAL METHOD}}\\\\
In this paper, we exclude photons, whose mass is zero, and molecules with velocities of the speed of light. For $m \neq 0$ and $v <c$, the Anderson-Witting model defined in eq. (29) can be rewritten as
\begin{eqnarray}
\frac{\partial f}{\partial t}+v^i \frac{\partial f}{\partial x^i}=&&\left(\gamma(u)(c,u_i)+\frac{q^\alpha}{ne+p}\right)(c,-v^i)^T \nonumber \\
 &&\frac{\left(f^{(0)}-f\right)}{c^2 \tau}
\end{eqnarray}
In general, the distribution function is $f=f(t,x^1,x^2,x^3,v^1,v^2,v^3)$, which has a one-to-one correspondence to $f=f(t,x^1,x^2,x^3,p^1,p^2,p^3)$. In this work we use $f=f(t,x^1,x^2,x^3,v^1,v^2,v^3)$ instead of $f=f(t,x^1,x^2,x^3,p^1,p^2,p^3)$. This transformation is shown in eq. (47) and can be done readily for the Marle model. To calculate the projected moments, we transform $d^3 \mathbf{p} /p^0$ into velocity space $d^3 \mathbf{v}$ as
\begin{eqnarray}
\frac{d^3 \mathbf{p}}{p^0}=J\left|\frac{\partial p^i}{\partial v^j}\right|/\left(m\gamma(v)c\right)=\frac{m^2\gamma(v)^4}{c} d^3 \mathbf{v}
\end{eqnarray}
From eq. (48), $N^\alpha$ can be rewritten as
\begin{eqnarray}
N^\alpha&=&c\int_{\mathscr{R}^3} p^\alpha f d^3\mathbf{p} /p^0\nonumber \\
&=& c \int_{\mathscr{V}^3} m \gamma(v) (c,v^i) f \frac{m^2 \gamma(v)^4}{c} d^3 \mathbf{v}\nonumber\\
&=&\int_{\mathscr{V}^3} m^3 \gamma(v)^5 (c,v^i) f d^3 \mathbf{v}.
\end{eqnarray}
$T^{\alpha\beta}$ can also be rewritten as
\begin{eqnarray}
T^{\alpha\beta}&=&c\int_{\mathscr{R}^3} p^\alpha p^\beta \frac{d^3 \mathbf{p}}{p^0} \nonumber \\
&=&\int_{\mathscr{V}^3} m^4 \gamma(v)^6 (c,v^i)(c,v^j) f d^3\mathbf{v}.
\end{eqnarray}
In eqs. (49) and (50), $\mathscr{V}^3$ is velocity space stretched by $\{\mathscr{V}^3 | (-c,-c,-c) \le (v^1,v^2,v^3)\le (c,c,c) \}$.\\
For convenience, non-dimensionalization is done as
\begin{eqnarray}
\tilde{n}&=&\frac{n}{n_\infty}, \  \tilde{v^i}=\frac{v_i}{c}, \ \tilde{u^i}=\frac{u^i}{c} \nonumber \\
\tilde{e}&=&\frac{e}{mc^2},\  \tilde{q^\alpha}=\frac{q^\alpha}{n_\infty mc^3} \nonumber \\
\tilde {x^i}&=&\frac{x^i}{L},\  \tilde{t}=\frac{t}{t_\infty},\  t_\infty=\frac{L}{c},
\end{eqnarray}
where $L$ is the representative length in the observer's frame.\\
With these non-dimensionalized quantities defined in eq. (51), the Maxwell-J\"uttner function in eq. (30) can be non-dimensionalized as
\begin{eqnarray}
\tilde f^{(0)}=\frac{(mc)^3}{n_\infty} f^{(0)}=\frac{\tilde n \zeta}{4\pi K_2(\zeta)} e^{-\zeta \gamma(\tilde u)\gamma(\tilde v)\left(1-\tilde u^i \tilde v^i \right)} \nonumber \\
\end{eqnarray}
To solve eq. (47), the second-order TVD (Total Variable Diminishing) scheme \cite{Yee} is used for the left hand side of eq. (47), and second order Runge-Kutta time integration is used for the time integration of eq. (47).\\
The left hand side of eq. (47) represents the propagation of molecules with velocity vector $(v^1,v^2,v^3)$ in physical space $(x^1,x^2,x^3)$. This formulation for molecular propagation in physical space does not involve relativistic effects. As a result, in body-fitted curvilinear coordinates $(\xi^1,\xi^2,\xi^3)$, molecules with a velocity vector $(v^1,v^2,v^3)$ in $(x^1,x^2,x^3)$ propagate with velocity $(v_{\xi}^1,v_{\xi}^2,v_{\xi}^3)$. Eq. (47) can therefore be written in body-fitted curvilinear coordinate $(\xi^1,\xi^2,\xi^3)$ as
\begin{eqnarray}
&&\frac{\partial \tilde f}{\partial t}+\frac{\partial v_\xi^i \tilde f}{\partial \xi^i}=\left(\gamma(u)(c,u_i)+\frac{q^\alpha}{ne+p}\right)(c,-v^i)^T \times \frac{\left(\tilde f^{(0)}-\tilde f\right)}{c^2 \tau}.\\
&& \tilde f=f/\tilde J, \nonumber
\end{eqnarray}
where $\tilde{J}$ is the Jacobian between $x^i$ and $\xi^i$.\\
The wall condition must also be considered. In this paper, complete diffusion at the wall is assumed. From conservation of the mass flux to the wall and by setting the $\xi^2$ axis as the normal vector to the plane element of the wall, we obtain the following relation
\begin{eqnarray}
f_w&=&f \ (v_\xi^2<0),\\
f_w/n_w&=&f^{(0)}(1,\theta_w,0)\ (v_\xi^2 \ge 0),\\
n_w&=&\frac{-\int_{v_\xi^2<0} v_\xi^2 f \gamma^5 d^3 \mathbf{v}}{\int_{v_\xi^2 \ge 0} v_\xi^2 f_w/n_w \gamma^5 d^3 \mathbf{v}}.
\end{eqnarray}
where $f_w$ is the distribution function on the wall. $n_w$ is the number density reflected from the wall and $\theta_w$ is the temperature of the wall.\\\\
\textsf{\textbf{V. RELATIVISTIC SHOCK LAYER PROBLEM}}\\\\
In this section we consider the formation of shock layers around circular cylinders. We model this problem using the Anderson-Witting, Marle, and energy-preserved Marle model, which is obtained by setting $\theta=\theta_E$ in eq. (38). For easier comprehension of physical conditions, for the observer's frame the absolute standard of rest is used as the hypothetical inertial frame. We use $(x,y,z)$ instead of $(x^1,x^2,x^3)$ and $(v_x,v_y,v_z)$ instead of $(v^1,v^2,v^3)$. The velocity corresponding to uniform flow is $u_x=0.5c,\ u_y=0,\ u_z=0$. Figure 2 shows a schematic view of the observer's frame and the flow field. The temperature of the uniform flow is $\theta_\infty=mc^2/45k$. Under these conditions, the Mach number of uniform flow calculated from eq. (33) is $2.689$. The temperature of the wall is $\theta_w=mc^2/30k$. From eq. (32) $\tau = 4\pi\sigma/L^2=10$. Molecules are all assumed to be monatomic hard-sphere molecules. For the numerical grid, $(v_x,v_y,v_z,x,y)=(64,64,64,81,60)$. Numerical tests indicate that this numerical grid provides accurate simulations. Figure 3 shows the number density profile, the velocity profile, and the temperature profile along the stagnation streamline. The shock thickness simulated with the Anderson-Witting model is thinner than that simulated with either the Marle or the energy-preserved Marle models. The shock thickness simulated with the Marle model is thinner than that simulated with the energy-preserved Marle model. The difference of the simulated shock thickness between Anderson-Witting and Marle models is discussed later in detail by considering the behavior of the negative velocity tail of the distribution function. The temperature simulated with the Marle model is lower than that simulated with the energy-preserved Marle model. This implies that the dynamic pressure $\varpi$ in eq. (46) is negative and that energy dissipation occurs.\\

Figure 4 shows profiles of heat flux $q^0$ and $q^x$ along the stagnation streamline. Both $q^0$ and $q^x$ have minimum values at the same point in the shock structure in all models and satisfy $q^\alpha U_\alpha=0$ in eq. (19). With the relation $q^\alpha U_\alpha=0$, $q^0 \simeq 0$ near the boundary layer is demonstrated by the fact that the flow velocity is about zero for the stagnation point. $q^x$ is approximated from eq. (28) by using the N.S.F law. From the spatial gradient of $\theta$ and $p$, $q^x_{NSF}$ approximated by the N.S.F. law is introduced from eq. (28). We define the heat flux by the gradient of the temperature as $q^{xt}_{NSF}=\lambda \nabla^\alpha \theta$ and by the gradient of the isotropic pressure as $q^{xp}_{NSF}=-\lambda \frac{\theta}{nh_E}\nabla p$. As a result, $q^x_{NSF}=q^{xt}_{NSF}+q^{xp}_{NSF}$. For results from the Anderson-Witting model, Figure 5 shows $q^x_{NSF}$ together with $q^x$ along the stagnation streamline on the left $y$ axis, and on the right $y$ axis Fig. 5 shows $|q^{xp}_{NSF}/q^{xt}_{NSF}|$ along the stagnation streamline, which is the ratio of absolute values of $q^{xt}_{NSF}$ and $q^{xp}_{NSF}$. The thermal conductivity $\lambda$ from the Anderson-Witting model, which is necessary for the calculation of $q^x_{NSF}$, $q^{xt}_{NSF}$ and $q^{xp}_{NSF}$, is given by \cite{Cercignani}. As shown in Fig. 5, $q^x \le q^x_{NSF}$ near $1.8 \le -X/R \le 3.2$ in the shock structure indicates significant effects by the terms from the Burnett equation. Near $-X/R \simeq 1.4$, $|q^{xp}_{NSF}/q^{xt}_{NSF}|$ exhibits a maximum value. As shown in Fig. 5, the heat flux calculated by the gradient of the isotropic pressure, $q^{xp}_{NSF}$, is nonnegligible for the calculated heat flux $q^x$ in this problem.\newline 

Figure 6 shows the profile of the dynamic pressure along the stagnation streamline. Under steady flow conditions, $\varpi=\frac{\eta \gamma(u)}{n}{u}_i \frac{\partial n}{\partial x_i}$ is obtained from eqs. (20) and (26). This indicates that the dynamic pressure depends on the product of the gradient of the number density and the flow velocity. As shown in Fig. 6, the dynamic pressure is negative in the shock structure and in the boundary layer, where both the gradient of the number density and flow velocity are positive. The dynamic pressure is positive in the sandwiched region defined by the shock structure and the boundary layer, $1.2 \le -X/R \le 2.0$. Assuming that the approximation by the N.S.F. law in eq. (26) is adequate to describe the behavior of the dynamic pressure in both the shock structure and in the boundary layer, eq. (26) indicates that $\eta$ is negative for all models in the shock structure and the boundary layer. We are continuing to investigate whether or not this negativity is caused by contributions from the Burnett terms \cite{Samojeden}. We do know, however, that the difference between the Marle and the energy-preserved Marle models in the bulk viscosity shown in Fig. 1 is not reflected in the difference of the profiles between these models, as shown in Fig. 6. However it is notable that the dynamic pressure has negative profiles in the boundary layer, where it might be less affected by the Burnett terms than in the shock structure.\newline

Figure 7 shows simulated distribution functions at selected points on the stagnation streamline for the Anderson-Witting and Marle models. In the shock structure, nonequilibrium conditions exist in the negative velocity tail, as shown in Figs. 7A and 7B. Near the positive peak of the dynamic pressure $-X/R=1.978$, Fig. 7C indicates that nonequilibrium conditions exist near the peak of the distribution function. Fig. 7D indicates that at $-X/R=1.246$, which is ahead of the boundary layer, that the distribution function represents a weak nonequilibrium condition at its peak. Fig. 7E indicates that at $-X/R=1.082$, which is the middle of the boundary layer, that the distribution function shifts slightly to the left from the equilibrium distribution function. Fig. 7F indicates that at $-X/R=1.0$, that the distribution function is not contiguous on both sides of $v_x/c=0$.\\

Finally, we consider the relaxation process of the distribution function in the shock structure for the Anderson-Witting and Marle models by focusing on the dynamics of the negative velocity tail of the distribution function. To explain the numerical results shown in Figs. 7A and 7B, the relaxation rate for both models is clarified. \newline

We introduce a new relaxation rate parameter, $\tilde \phi_{AW}$, which is obtained by removing $\frac{q^\alpha}{ne+p}$ and $\tau$ from the relaxation rate $\frac{\left(\gamma(u)(1,\tilde u_i)+\frac{\tilde q^\alpha}{\tilde n\tilde e+\tilde p}\right)(1,-\tilde v^i)^T}{\tau}$ of the right hand side of eq. (47) as
\begin{eqnarray}
\tilde\phi_{AW}=\frac{1-\tilde u_i \tilde v_i}{\left(1-\tilde u^2\right)^{1/2}}.
\end{eqnarray}
We restrict ourselves to $\tilde\phi_{AW}$ along the stagnation streamline, for which $u_i=(u_x,0,0)$ and $v_i=(v_x,v_y,v_z)$ gives $\phi_{AW}$ as the specific case of $\tilde\phi_{AW}$,
\begin{eqnarray}
\phi_{AW}=\frac{1-\tilde u_x \tilde v_x}{\left(1-\tilde u_x^2\right)^{1/2}}.
\end{eqnarray}
For comparison, the relaxation rate parameter for the Marle model, $\phi_M$, is obtained by rewriting eq. (38) into $\frac{\partial f}{\partial t}+v_i \frac{\partial f}{\partial x^i}=\frac{1}{\gamma(v)\tau}\left(f^{(0)}-f\right)$ and replacing $\phi_M=\frac{1}{\gamma(v)}$ with $v_y=v_z=0$ from the right-hand side of eq. (47). Finally $\phi_M$ can be expressed as
\begin{eqnarray}
\phi_M=\left(1-\tilde v_x^2\right)^{1/2} \ge \left(1-\tilde v^2 \right).
\end{eqnarray}
Figure 8 shows $\phi_{AW}$ versus $v_x$ for various values of $u_x$, and also shows $\phi_M$ versus $v_x$. When the gas is at rest, $u_x=0$, $\phi_{AW}=1$ for all values of $v_x$. Generally $\phi_{AW}$ is the envelope for the half circle of $\phi_M$, which contacts the half circle of $\phi_M$ at $v_x=u_x$. Because the absolute value of $u_x$ is approximately equal to the speed of light, $\phi_{AW}$ is approximately infinity, except for when $|u_x|=c$, in which case $\phi_{AW}$ approaches zero as $|u_x|\rightarrow c$. On the other hand, $\phi_M$ decreases as the absolute value of $v_x$ increases, regardless of the magnitude of $u_x$. In the shock layer, the extent of nonequilibrium in the negative velocity tail determines the shock thickness, because nonequilibrium first appears in the negative velocity tail. Figure 7 indicates more rapid relaxation in the negative tail in the Anderson-Witting model than in the Marle model. The more rapid relaxation yields lower population {\sf Lower population of what?} in the negative velocity tail. Consequently the shock thickness simulated with the Anderson-Witting model is thinner than that simulated with the Marle model.\\\\
\textsf{\textbf{VI. CONCLUSIONS}}\\\\
In this paper, we used two different relativistic Boltzmann-kinetic equations, the Anderson-Witting and Marle models, to numerically solve a shock-layer problem. The simulated heat flux has similar characteristics to that approximated by using the N.S.F. law. The heat flux calculated by using the gradient of the isotropic pressure represents a nonnegligible component of the total heat flux calculated for this problem. On the other hand, the simulated behavior of the dynamic pressure shows opposite tendencies to that approximated by using the N.S.F. law. It is important to determine whether or not this discrepancy between the N.S.F law and the simulated dynamic pressure is caused by the contribution of terms above the Burnett equation. The relaxation rate of the distribution function by using the Anderson-Witting model depends on both the flow velocity and on the molecular velocity. On the other hand, the relaxation rate simulated with the Marle model depends only on the molecular velocity. This difference of relaxation rate between these two models is manifested as a difference of the thickness of the shock structure.\\\\
\textsf{\textbf{APPENDIX}}\\\\
\renewcommand{\theequation}{A.\arabic{equation}}
\setcounter{equation}{0} $\psi(\zeta_E)$ in eq. (41) is given by
\begin{eqnarray}
\psi(\zeta_E)=\tau mkc^2 \frac{20G_E+3\zeta_E-13G_E^2\zeta_E^2-2G_E^2\zeta_E^2-2G_E\zeta_E^2+2G_E^3\zeta_E^3}{\zeta_E C_v(\zeta_E)},
\end{eqnarray}
where $C_v(\zeta_E)$ is the constant-volume specific heat given by $C_v(\zeta_E)=k\left(\zeta_E^2+5G_E\zeta_E-G_E^2\zeta_E^2-1\right)$.\\
The correct bulk viscosity for $\theta \neq \theta_E$ is given by \cite{Cercignani}
\begin{eqnarray}
\eta(\zeta_E)=\frac{\tau p_Ek^2}{3} \frac{\left(20G_E+3\zeta_E-13G_E^2\zeta_E-2G_E\zeta_E^2+2G_E^3\zeta_E^2\right)\left(4-\zeta_E^2-5G_E\zeta_E+G_E^2\zeta_E^2 \right)}{C_v(\zeta_E)^2}
\end{eqnarray}
The bulk viscosity for $\theta=\theta_E$ is given by
\begin{eqnarray}
\tilde\eta(\zeta)=\frac{\tau pk^2}{3} \frac{\left(20G+3\zeta-13G^2\zeta-2G\zeta^2+2G^3\zeta^2\right)\left(1-\zeta^2-5G\zeta+G^2\zeta^2 \right)}{C_v(\zeta)^2}
\end{eqnarray}

\newpage
\hspace{-1.5em}\textsf{\textbf{FIGURE CAPTIONS}}\\\\
FIG. 1: Bulk viscosity for the Marle model for $\theta=\theta_E$ and $\theta \neq \theta_E$.\\
FIG. 2: Schematic view of observer's frame and flow field.\\
FIG. 3: Number density, temperature, and velocity profiles along the stagnation streamline for the A.W. (Anderson-Witting), Marle, and energy-preserved Marle models $(\theta=\theta_E)$.\\
FIG. 4: Heat flux ($q^0$ and $q^x$) profiles along the stagnation streamline for the A.W. (Anderson-Witting), Marle, and energy-preserved Marle models $(\theta=\theta_E)$. \\
FIG. 5: Heat flux $q^x_{NSF}$ approximated by using the N.S.F. law and $|q^{xp}_{NSF}/q^{xt}_{NSF}|$ along the stagnation streamline for the A.W. (Anderson-Witting) model. \newline
FIG. 6: Dynamic pressure profiles along the stagnation streamline for the A.W. (Anderson-Witting), Marle, and energy-preserved Marle models $(\theta=\theta_E)$. \\
FIG. 7 (A)-(E): Distribution functions and equilibrium distribution functions at selected points (A) $-X/R=4.965$, (B) $-X/R=3.218$, (C) $-X/R=1.978$, (D) $-X/R=1.246$, (E) $-X/R=1.082$, and (F) $-X/R=1.00$ (Wall) on the stagnation streamline for the A.W. (Anderson-Witting) and Marle models. \\
FIG. 8: Relaxation rate parameters $\phi_{M}$ and $\phi_{AW}$ for various values of $u_x/c$ versus $v_x/c$.
\newpage
\begin{center}
\includegraphics[width=1.\linewidth]{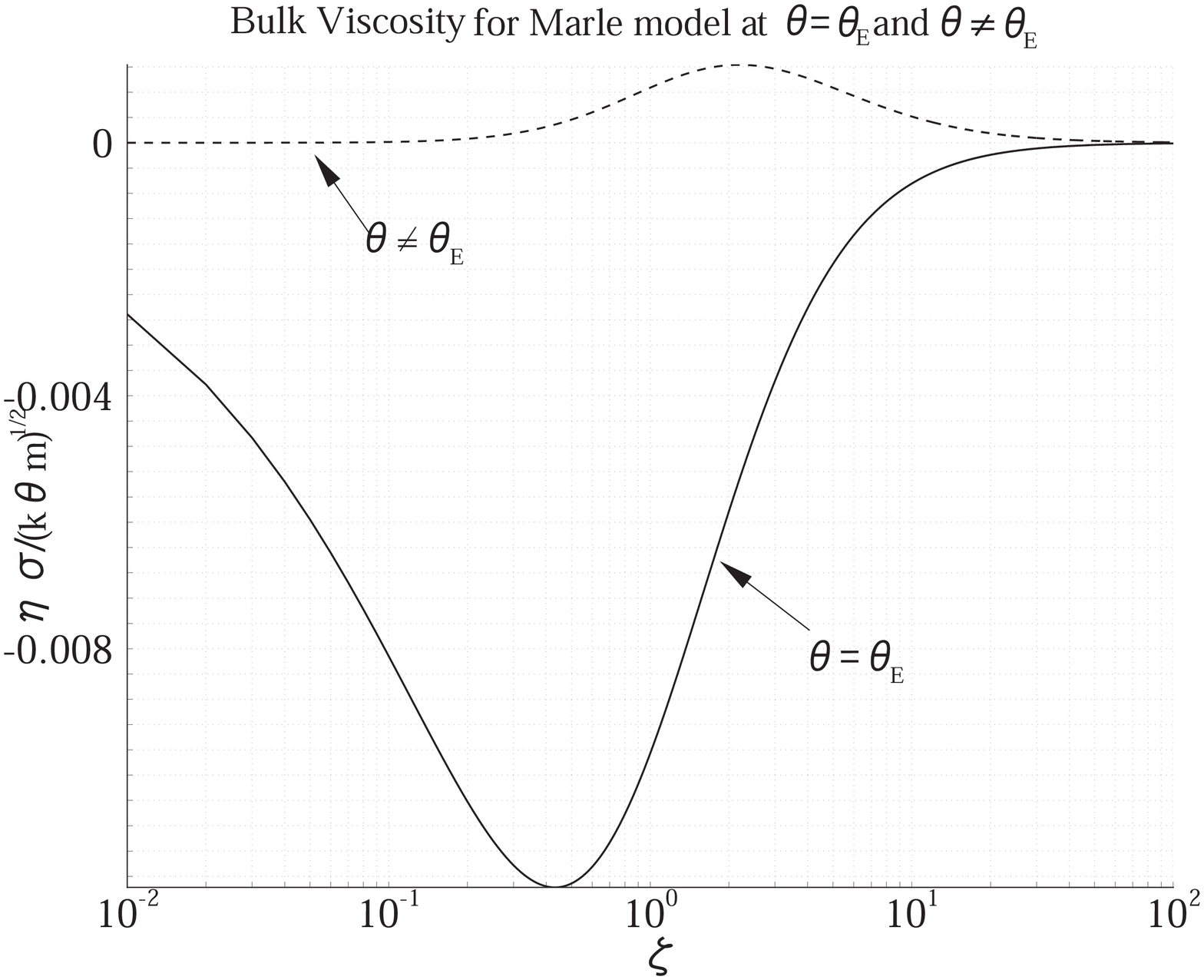} \\
\footnotesize{FIG. 1 Bulk viscosity for the Marle model at $\theta=\theta_E$ and $\theta \neq \theta_E$.}\\
\footnotesize{($\sigma$ is the collision cross section for hard-sphere molecules) }
\end{center}
\begin{center}
\includegraphics[width=.9\linewidth]{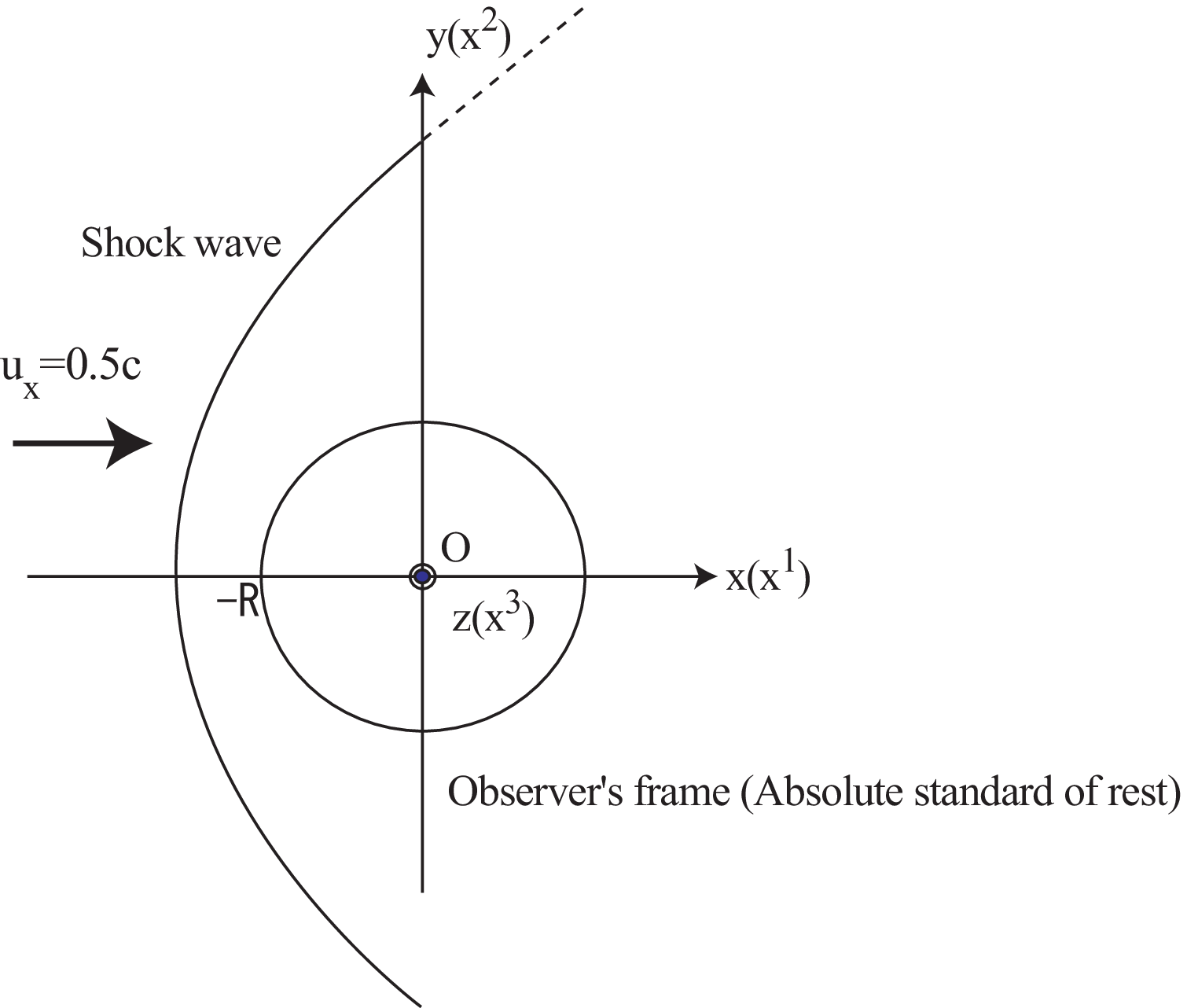} \\
\footnotesize{FIG. 2 Schematic view of observer's frame and flow field.}
\end{center}
\begin{center}
\includegraphics[width=1.\linewidth]{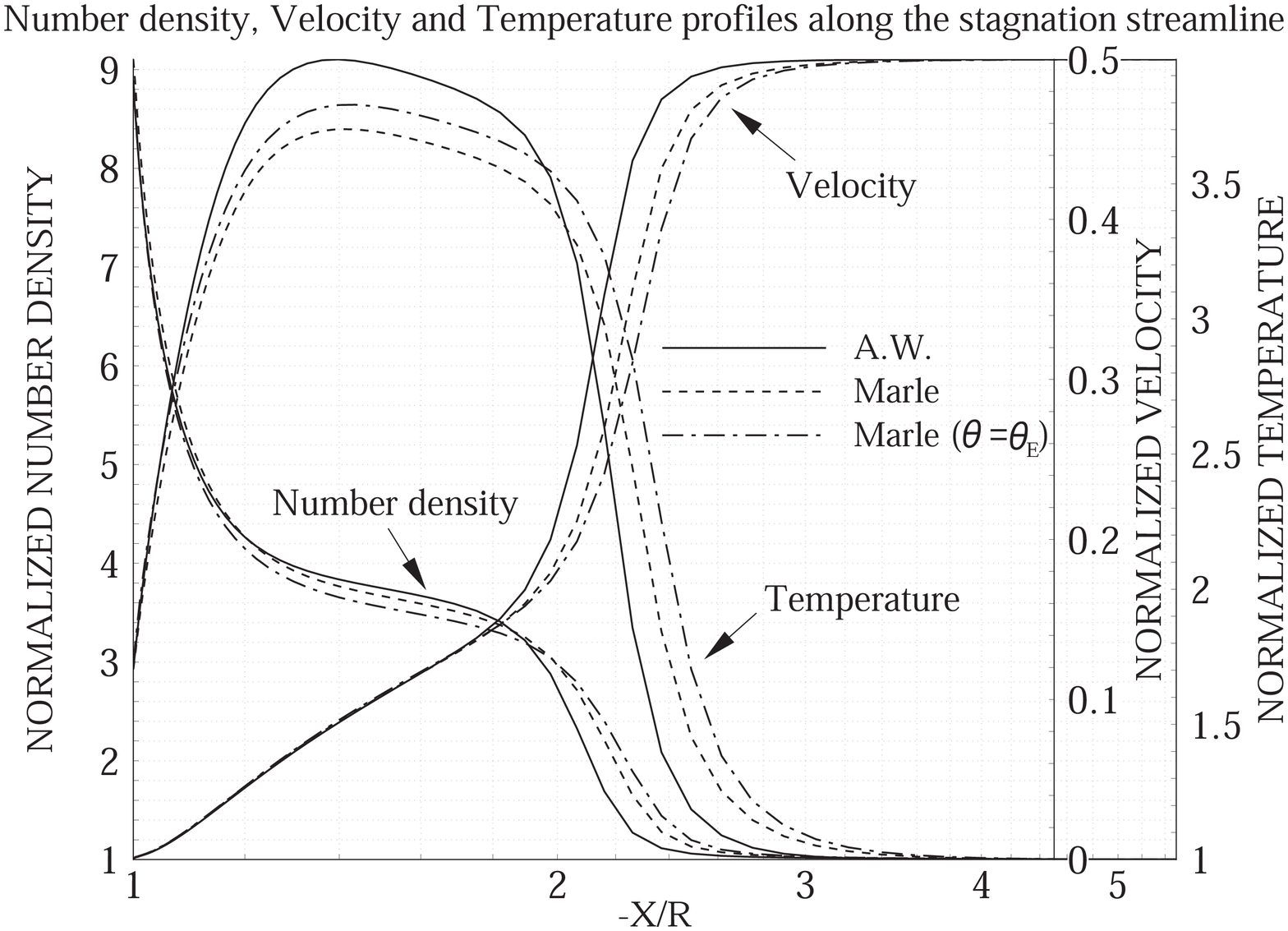} \\
\footnotesize{FIG. 3 Number density, temperature, and velocity profiles along the stagnation streamline.}
\end{center}
\begin{center}
\includegraphics[width=1.\linewidth]{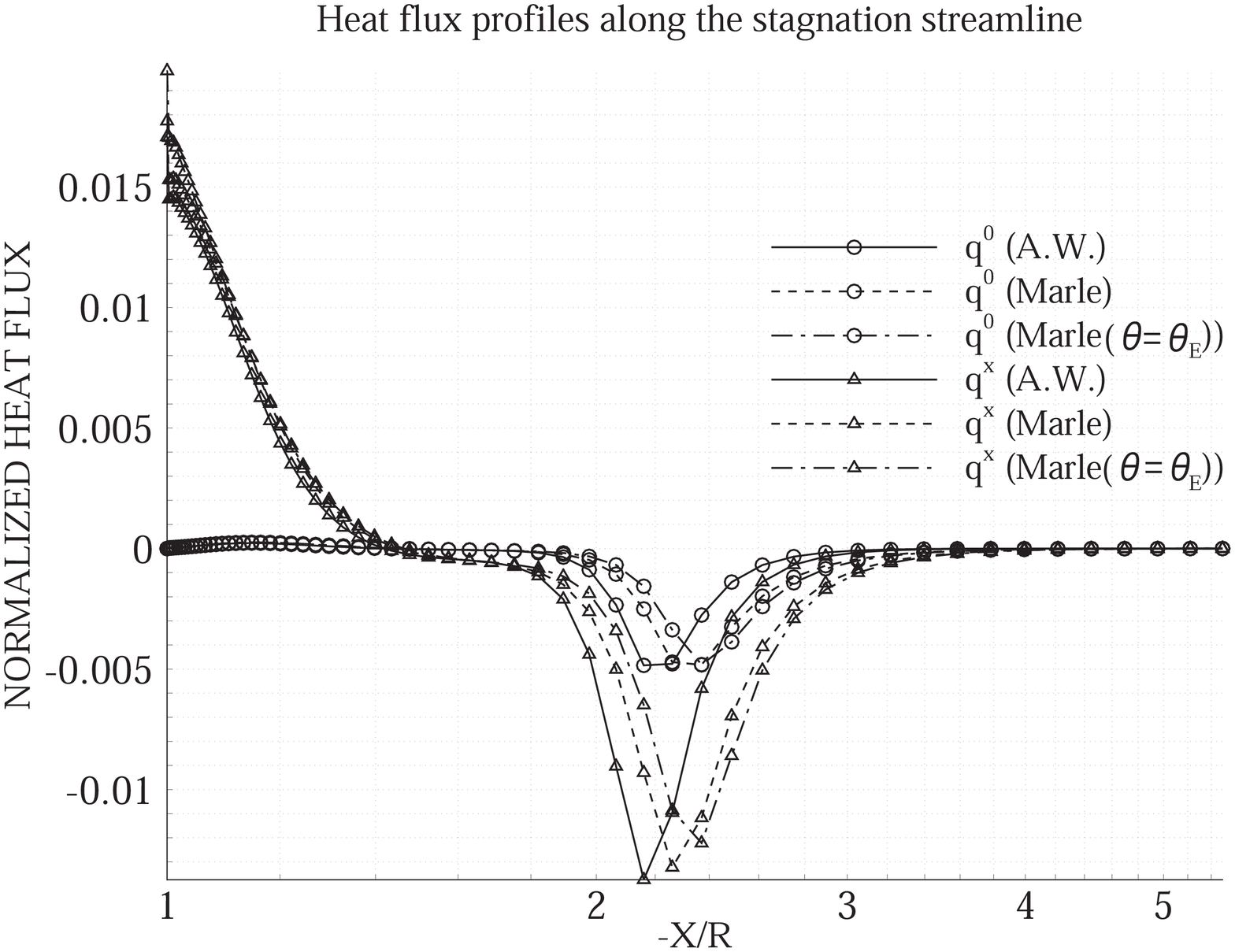}\\
\footnotesize{FIG. 4 Heat flux ($q^0$ and $q^x$) profiles along the stagnation streamline.}
\end{center}
\begin{center}
\includegraphics[width=1.\linewidth]{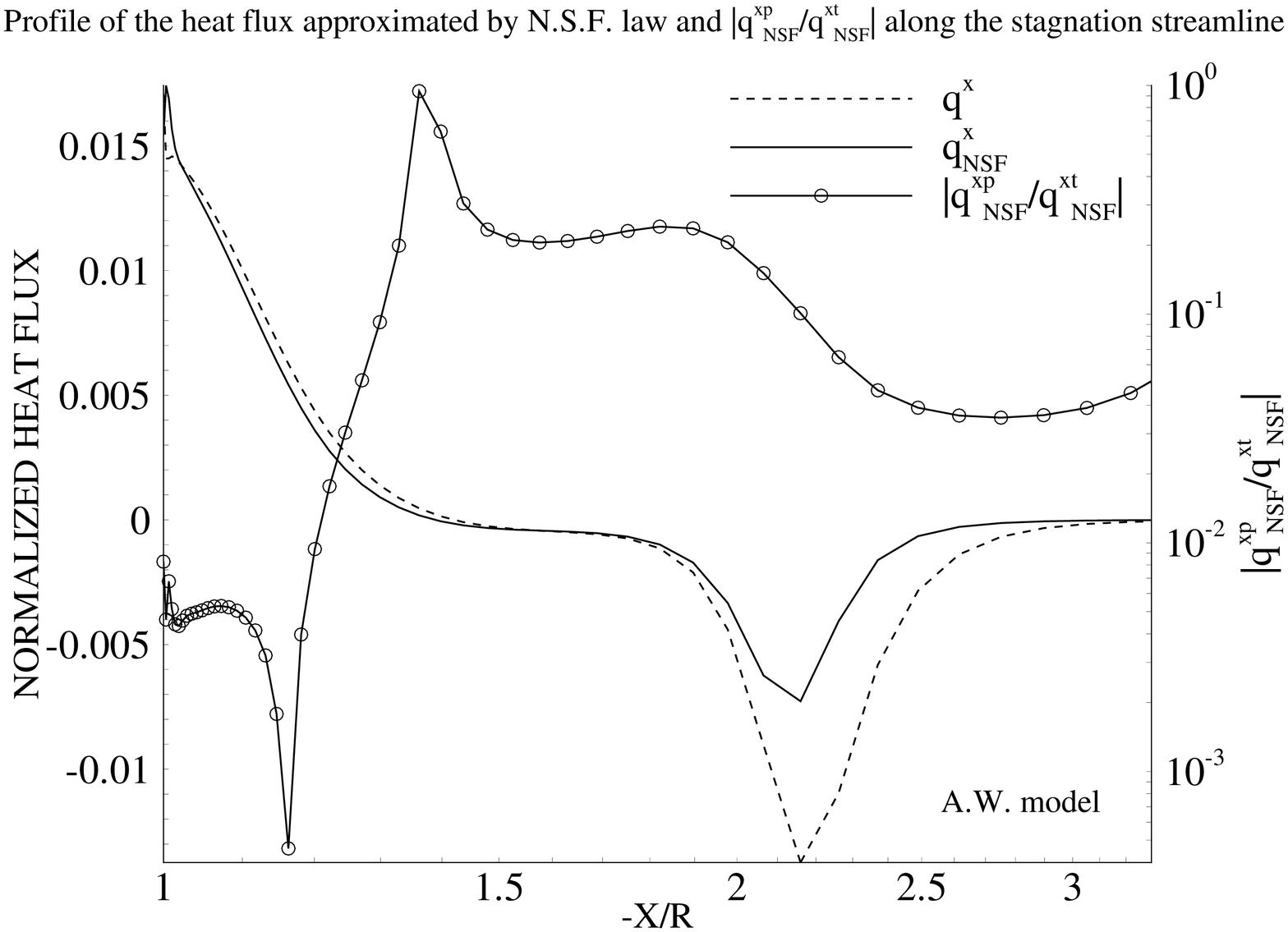}\\
\footnotesize{FIG. 5 Heat flux $q^x_{NSF}$ approximated by using the N.S.F. law and $|q^{xp}_{NSF}/q^{xt}_{NSF}|$ along the stagnation streamline for the A.W. model.}
\end{center}
\begin{center}
\includegraphics[width=1.\linewidth]{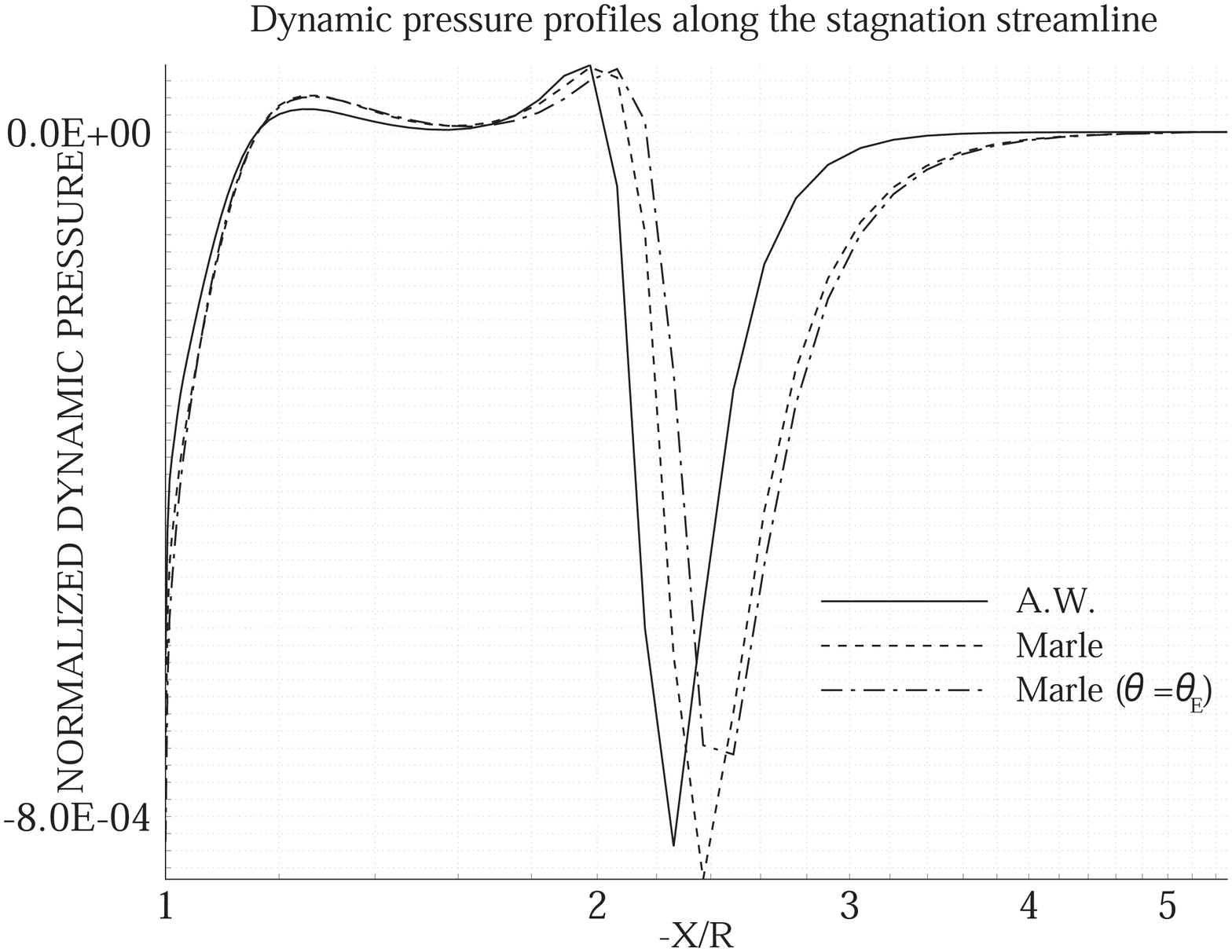}\\
\footnotesize{FIG. 6 Dynamic pressure profiles along the stagnation streamline.}
\end{center}
\begin{center}
\includegraphics[width=1.\linewidth]{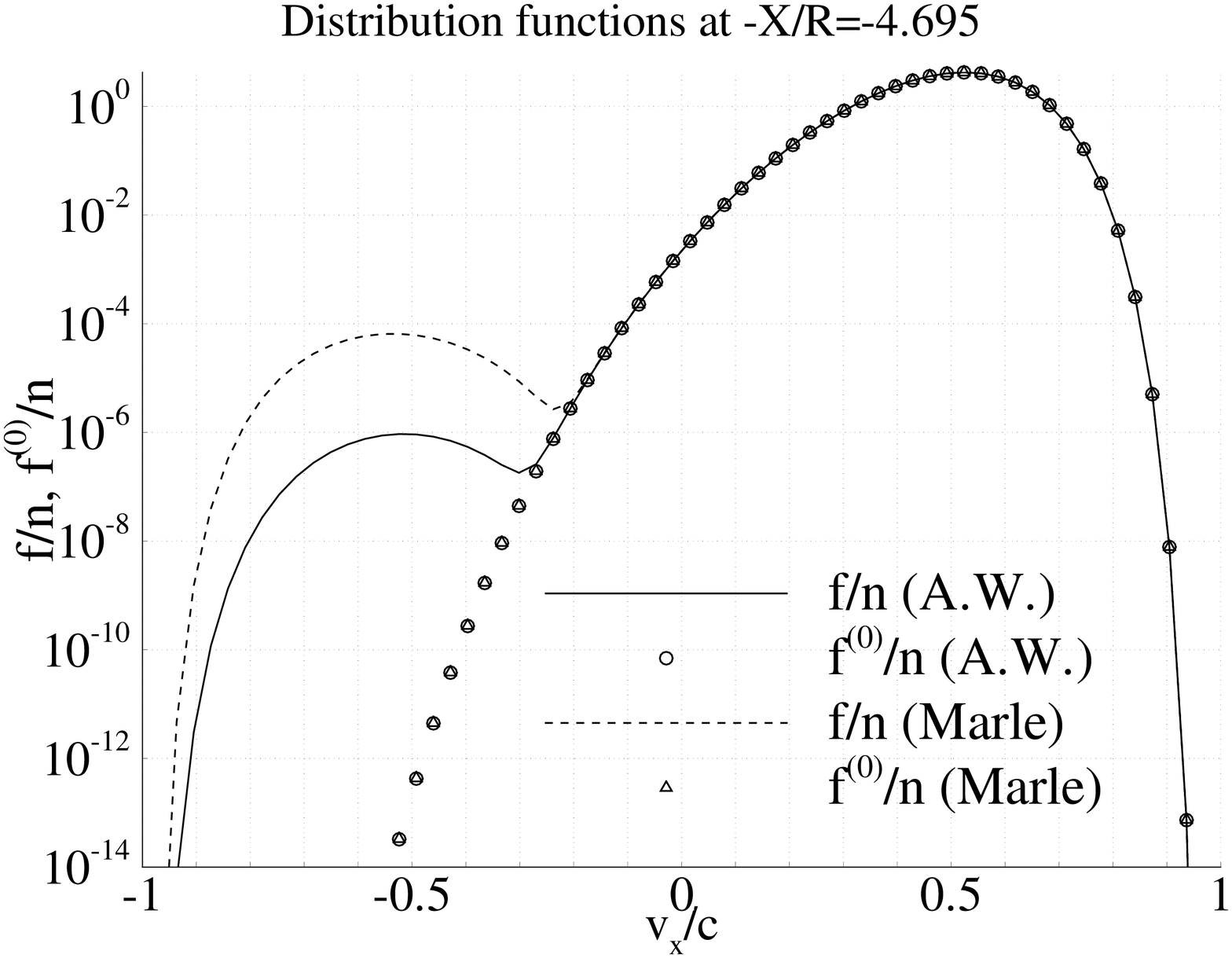} \\
\footnotesize{FIG. 7 (A): $-X/R=4.695$}
\end{center}
\begin{center}
\includegraphics[width=1.\linewidth]{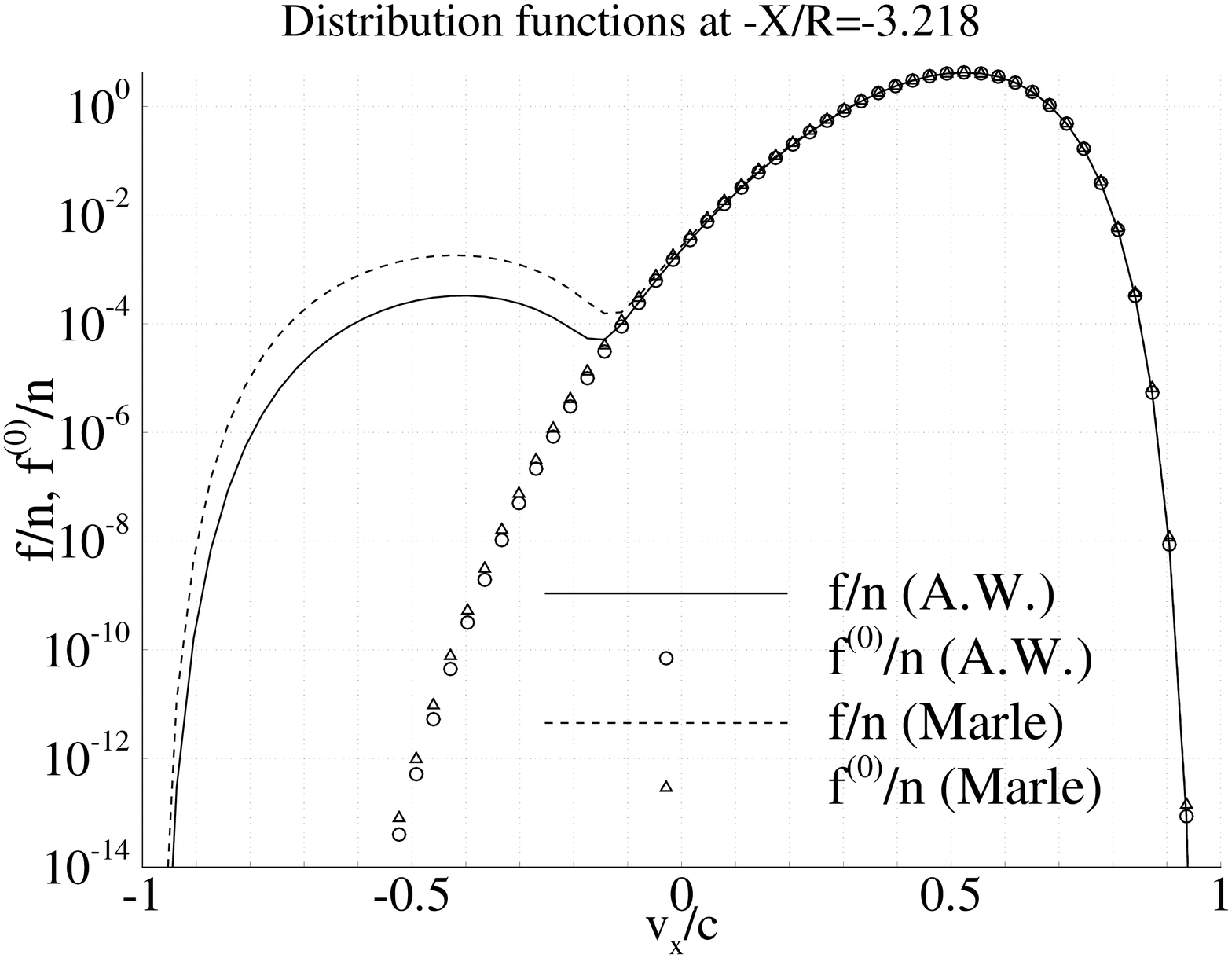} \\
\footnotesize{FIG. 7 (B): $-X/R=3.218$}
\end{center}
\begin{center}
\includegraphics[width=1.\linewidth]{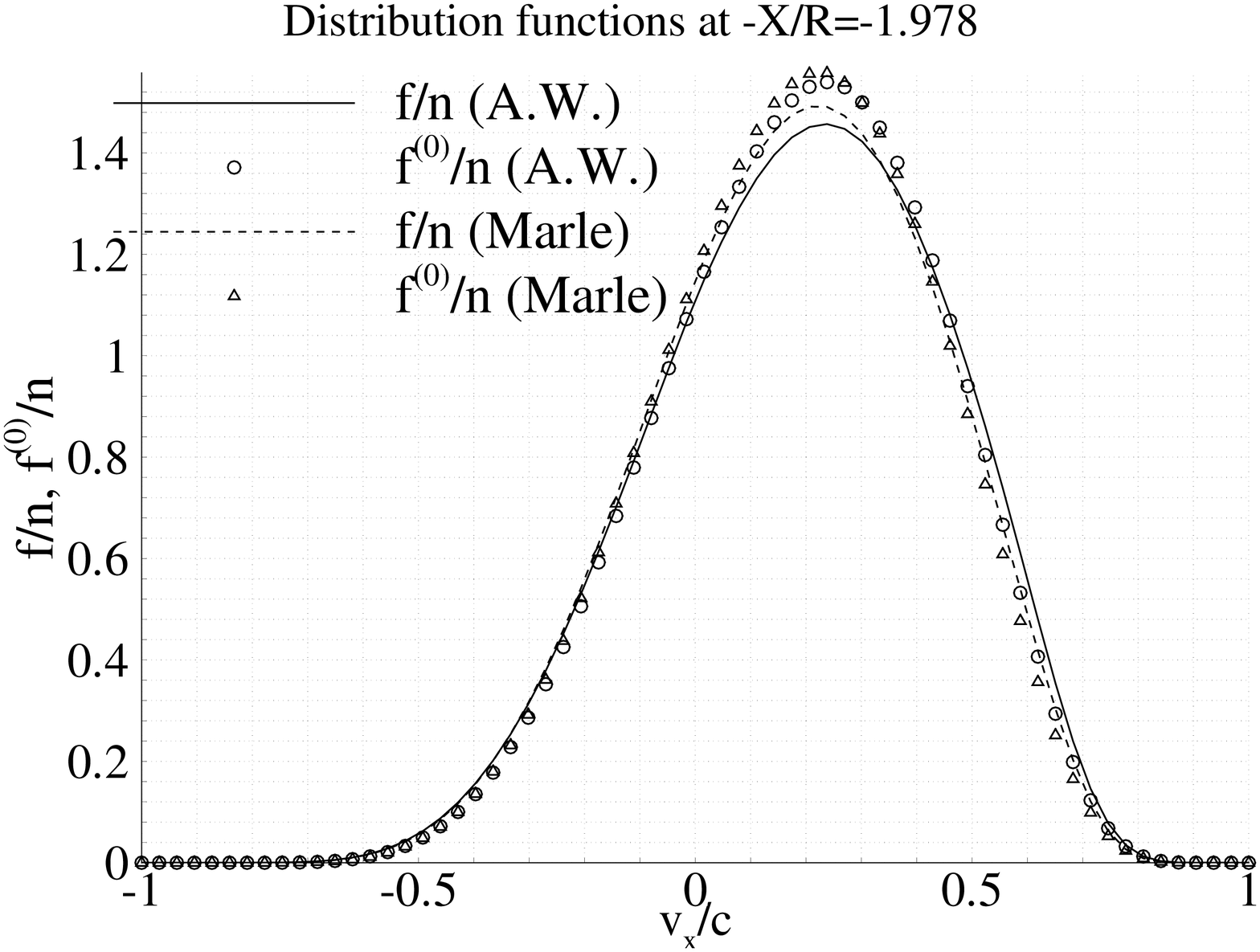} \\
\footnotesize{FIG. 7 (C): $-X/R=1.978$}
\end{center}
\begin{center}
\includegraphics[width=1.\linewidth]{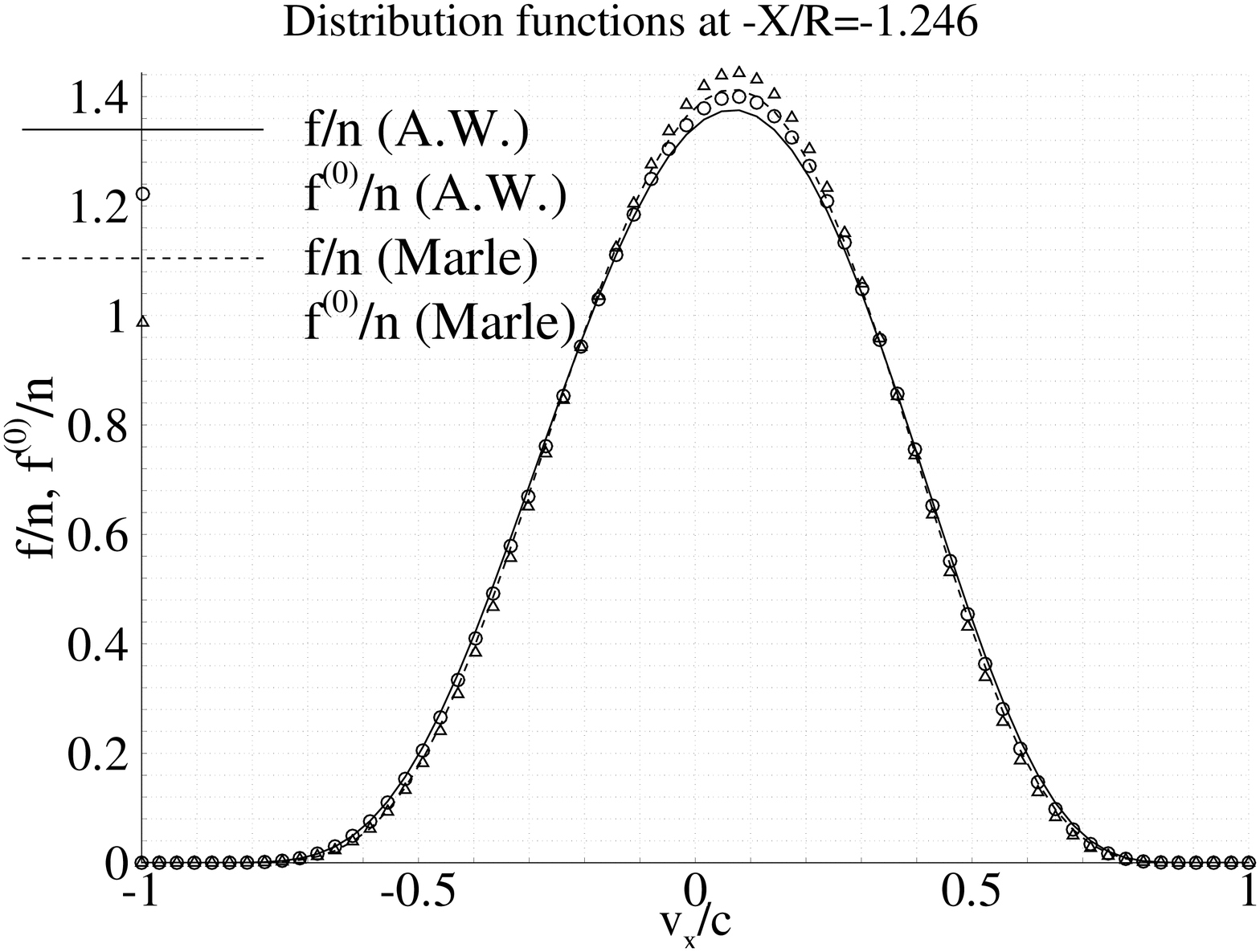} \\
\footnotesize{FIG. 7 (D): $-X/R=1.246$}
\end{center}
\begin{center}
\includegraphics[width=1.\linewidth]{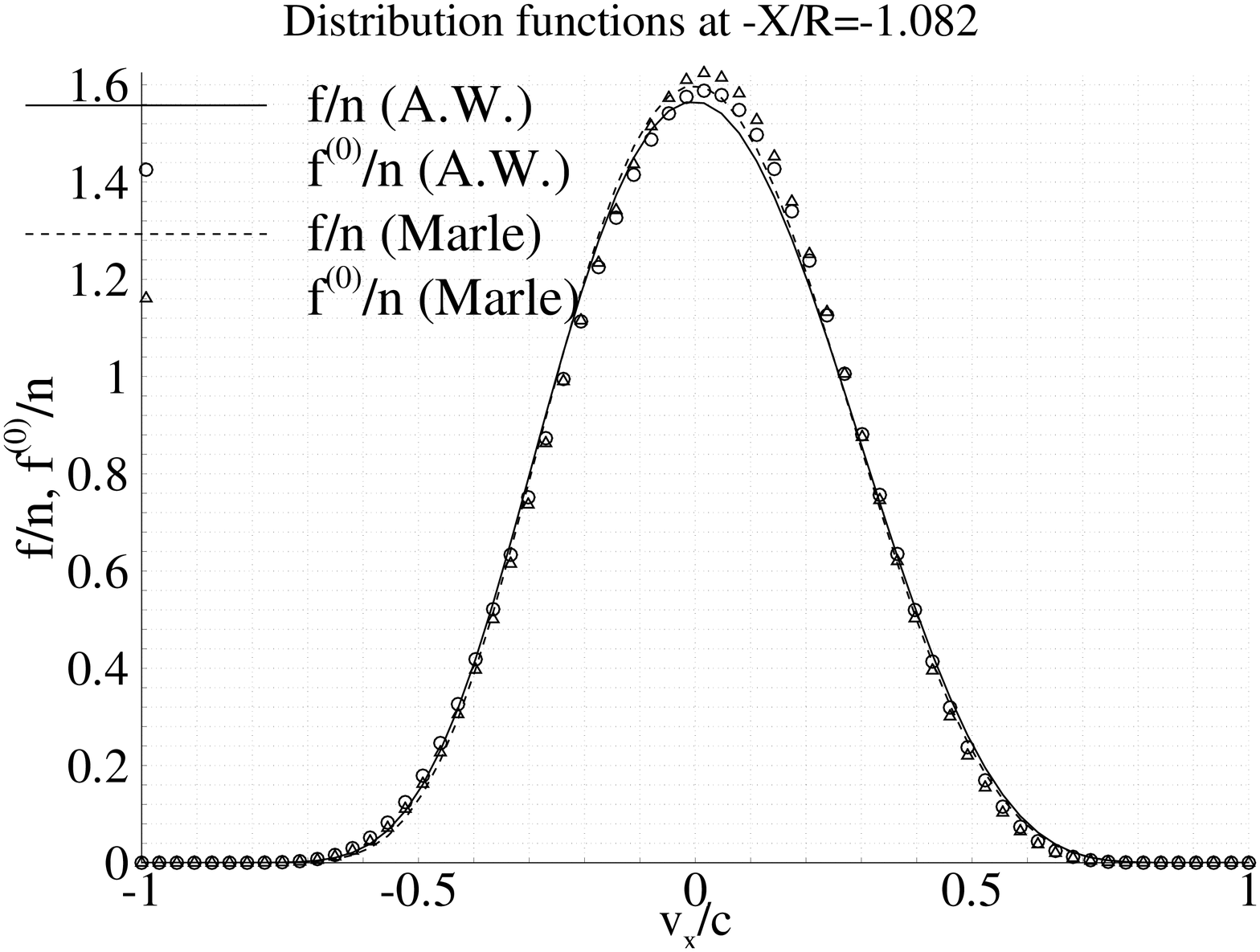} \\
\footnotesize{FIG. 7 (E): $-X/R=1.082$}
\end{center}
\begin{center}
\includegraphics[width=1.\linewidth]{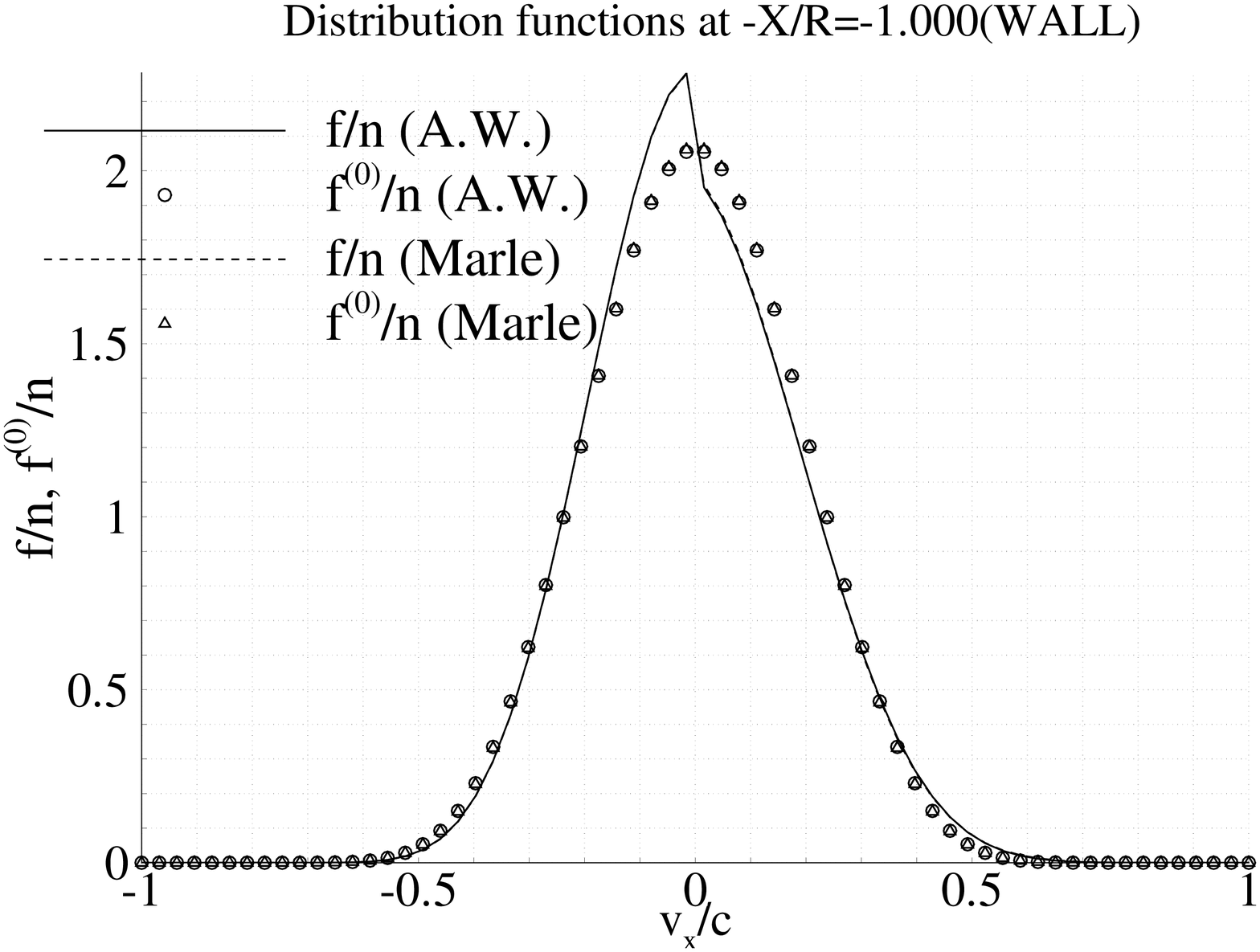} \\
\footnotesize{FIG. 7 (F): $-X/R=1.00$ (Wall)}
\end{center}
\begin{center}
\includegraphics[width=1.\linewidth]{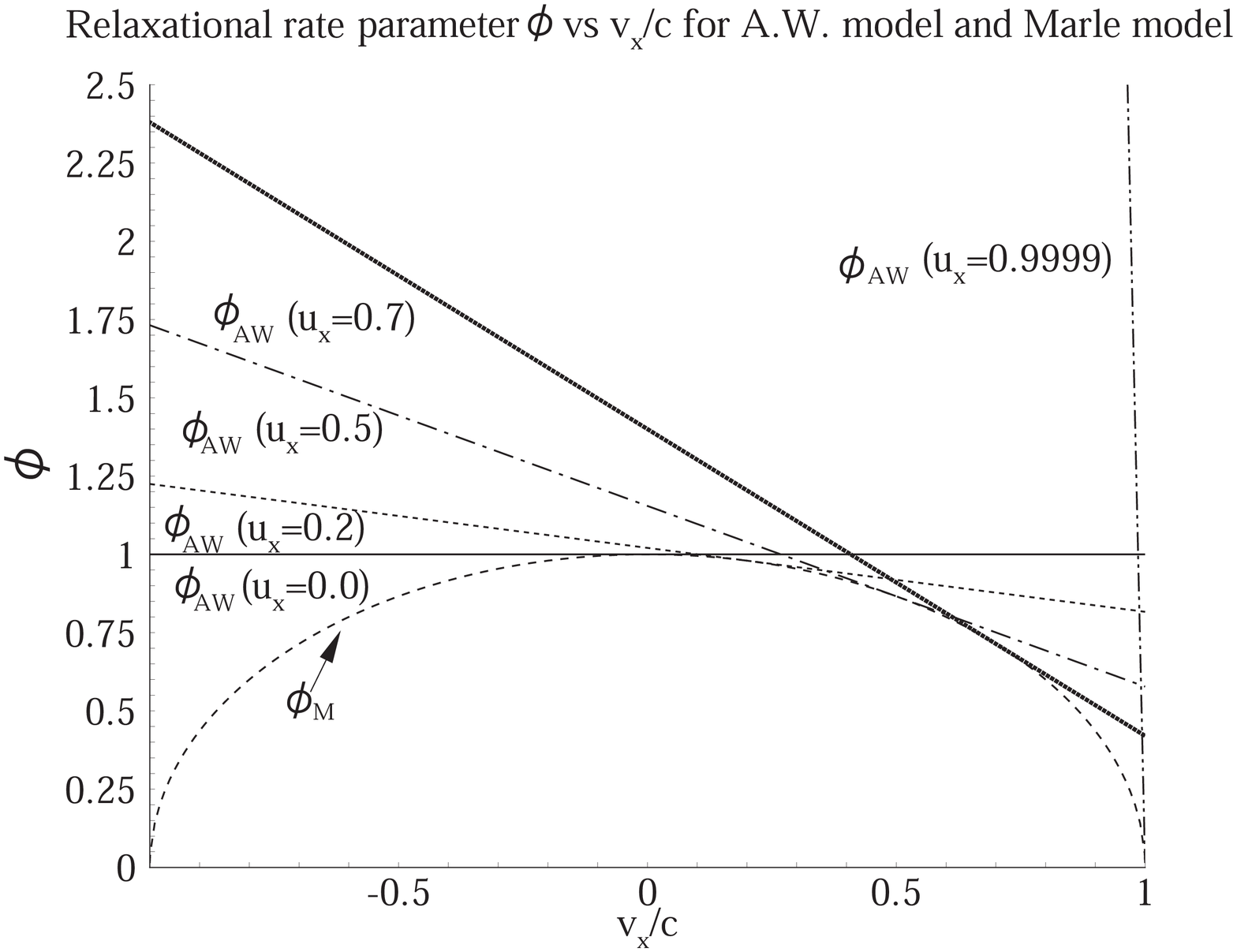} \\
\footnotesize{FIG. 8 Relaxation rate parameters $\phi_{M}$ and $\phi_{AW}$ for various values of $u_x/c$ versus $v_x/c$.}
\end{center}
\end{document}